# Dissipation in Quasi One-Dimensional Superconducting Single-Crystal Sn Nanowires

Ming-Liang Tian, Jin-Guo Wang, James S. Kurtz, Ying Liu, Theresa S. Mayer, Thomas E. Mallouk and M. H. W. Chan

*The Materials Research Institute and The Center for Nanoscale Science, the Pennsylvania State University, University Park, Pennsylvania 16802-6300.*

Electrical transport measurements were made on single-crystal Sn nanowires to understand the intrinsic dissipation mechanisms of a one-dimensional superconductor. While the resistance of wires of diameter larger than 70 nm drops precipitately to zero at $T_c$ near 3.7 K, a residual resistive tail extending down to low temperature is found for wires with diameters of 20 and 40 nm. As a function of temperature, the logarithm of the residual resistance appears as two linear sections, one within a few tenths of a degree below $T_c$ and the other extending down to at least 0.47 K, the minimum temperature of the measurements. The residual resistance is found to be ohmic at all temperatures below $T_c$ of Sn. These findings are suggestive of a thermally activated phase slip process near $T_c$ and quantum fluctuation-induced phase slip process in the low temperature regime. When the excitation current exceeds a critical value, the voltage-current (V-I) curves show a series of discrete steps in approaching the normal state. These steps cannot be fully understood with the classical Skocpol-Beasley-Tinkham phase slip center model (PSC), but can be qualitatively accounted for partly by the PSC model modified by Michotte *et al.*

When the diameter of a superconducting wire is smaller than the phase coherence length, $\xi$ (T), its behavior is expected to deviate from that of bulk and crosses over towards that expected of a quasi one-dimensional (1d) system. In spite of extensive experimental studies over the last three decades, there are still controversies on what are the expected properties of a 1d superconductor[1-6]. A major reason for the uncertainties is the variety of microstructure and morphology of the samples used in the experiments. Indeed, contrasting results are found in granular[1], polycrystalline[2] and amorphous wires[3-6] fabricated by sputtering or evaporating techniques. Measurements on single-crystal nanowires with uniform diameter would be ideal to single out the effect of 1d confinement. To date, such measurements were carried out only on crystalline superconducting whiskers with diameters ranging from 0.1 to 0.8 $\mu$m[7-10]. At such a length scale, 1d behavior is unlikely to be evident except at temperatures very close to $T_c$.

In this paper we present a systematic study of the transport properties of single-crystal cylindrical tin (Sn) nanowires with diameters between 20 and 100 nm. We chose tin in our study because the coherence length of bulk tin is relatively long ($\xi$ (0)~200 nm) and single-crystal nanowires of uniform diameter can be consistently prepared by a simple template-assembly technique[11]. Our results show a clear crossover from bulk-like to probably quasi 1d-like behavior when the diameter of the wires is reduced to 40 nm (5 times smaller than the bulk coherence length). Two different dissipative processes, i.e., thermally activated phase slip (TAPS) close to $T_c$ and quantum phase-slip (QPS) at temperatures far below $T_c$ are clearly observed for the wires of 20 and 40 nm in diameters. Current-induced multiple voltage steps in the voltage-current (V-I) characteristics were also observed in these single-crystal nanowires over a wide temperature range below $T_c$.

Tin nanowires were fabricated by electrodepositing tin into a porous membrane at room temperature[11]. The electrolyte was 0.1 M $SnSO_4$ aqueous solution with 2% gelatin by weight and the pH value was adjusted near 1 with concentrated $H_2SO_4$. Pure bulk tin wire was used as the positive electrode, and a Au film evaporated on the one side of the membrane worked as the negative electrode. The depositing voltage between two electrodes is about -80 mV. The wires with diameters of 40, 60, 70 and 100 nm and a length of about 6 $\mu$m were synthesized with commercial polycarbonate membranes (PCM) (Structure Probe, Inc(SPI),USA), while the 20 nm wires with a length of ~30 $\mu$m were made using "home-made" anodic alumina membrane (AAM). The images of the surface and the cross-section of the AAM are, respectively, shown in Fig.1(a) and 1(b), characterized by field emission scanning electron microscopy (FESEM). The 1d channels of the pores are found to be aligned almost parallel to each other and perpendicular to the surface of the membrane without interconnecting channels between the adjacent pores. The pore density of 20 nm AAM is about $4\times10^{10}$ pores/$cm^2$. For PCM membranes, the pore density is about $6\times10^8$ pores/$cm^2$, two orders lower than that of AAM. The pores in PCM are randomly distributed[12]. Fig.1(c) and 1(d), respectively, show the images of the resulting nanowires fabricated with PCM, characterized by transmission electron microscopy (TEM) and by electron diffraction (ED) measurements. All the nanowires randomly selected for the TEM study showed single-crystal structure along the entire length of the wire, and almost 90% of them showed a preferred [100] crystallographic direction of tetragonal $\beta$-tin. The diameter of an individual wire was found to be uniform along its length, but the diameters from wire to wire varies by about ±5 nm,

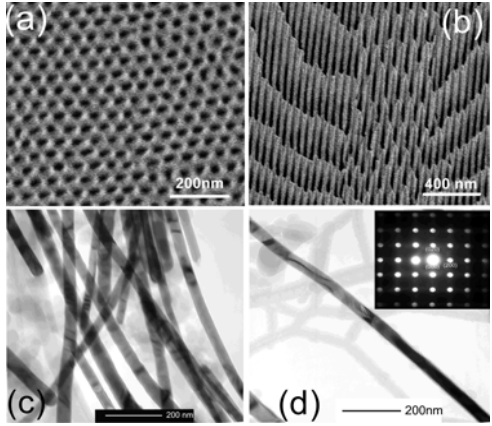

Fig.1, Panels (a) and (b), respectively, show the FESEM images of the surface and cross-section of "home-made" AAM membranes. (c) and (d), respectively, show the TEM images of the randomly distributed free-standing Sn nanowires fabricated with PCM, and the electron diffraction pattern of the individual 40 nm wire showing [100] orientation.

reflecting the dispersion of the pore diameter of the membrane.

Transport measurements were carried out with a Physical Properties Measurement System (PPMS) (Quantum Design Inc.), equipped with a He-3 cryostat and a superconducting magnet. Sn nanowires released from the membrane with diameter less than 40 nm are found to be unstable in shape at room temperature[13], i.e., in a matter of hours after released from the membrane, the wires are found to develop regions of thick nodules and narrow necks along the length of the wire. We think this is due to the fact that the wires are undergoing coarsening transformation towards their thermodynamic equilibrium, i.e., spherical, shape. Wires released from the membrane also develop oxide layer on the surface. These complications make standard four-lead measurement on thin individual Sn wire difficult. In this experiment, transport measurements are made on wires embedded in the membrane, the voltage and current leads to the wires are made by mechanically squeezing high-purity (99.999%) bulk Sn wires of 0.5 mm diameter onto the two sides of the membrane. Before the Sn leads were attached, the Au film pre-evaporated on the membrane for electrodeposition was removed either under $N_2$ protection atmosphere or with the membrane immersed in pure ethanol solution. A schematic of our experimental arrangement is shown as an inset of Fig.2 (a). This arrangement is very similar to the technique used for transport measurement of superconducting whiskers[7-9]. The whiskers were held by two electrically isolated superconducting blocks of Wood's metal, enabling two-probe measurements. There are reports of two-lead transport measurements on Au-capped polycrystalline Pb and Sn nanowires also embedded inside the membrane[12,14]. The wires are also fabricated by electrodeposition technique. The contact and lead resistance due to the gold caps cannot be subtracted from the data in those measurements. In our arrangements, it appears the procedure of squeezing the bulk Sn leads to the nanowire arrays can reliably break through the possible oxide layers and join the bulk Sn directly to the nanowires. The series resistance of bulk Sn leads and the contact resistance between the leads and the nanowire arrays for all samples we have studied, as we shall show below, were found to be negligibly small at temperatures below $T_c$ of bulk Sn. Therefore any features in the resistance below 3.7 K can be attributed to the nanowires.

In a separate measurement, the resistivity of a single 70 nm thick Sn wire released from membrane and attached to conducting leads fabricated by e-beam technique[15] was determined to be 7.6 $\mu\Omega$.cm at 5.0 K just above the transition temperature. Since the fabrication process is identical, it is reasonable to assume this wire has similar resistivity value as the wires still embedded in the membrane. This assumption allows us to estimate the number of wires in a membrane making contact to the bulk Sn leads from the measured resistance at 5 K. The estimated numbers of wires for the 20, 40, 60, 70 and 100 nm samples are, respectively, 18, 1, 8, 15 and 53. Based on the uncertainties of the length and diameter of the wires and possible variation in the resistivity value, the numbers listed above for the 20, 60, 70 and 100 nm samples are likely to be correct to within 25%. The estimate of a single wire for the 40 nm sample, however, must be accurate as the stated uncertainties in resistivity and wire dimensions do not allow any conclusion other than that of a single wire. We note that none of our conclusions reported in this paper depend on the number of wires making contact to the bulk Sn leads. The results we found on the single 40 nm wire does not show any unusual features as compared to those of the multi-wire arrays and, taken together, the data of all the samples display a consistent trend with decreasing wire diameter.

In Fig.2 (a), the solid line shows the R(T) curve of 20 nm Sn wire array (18 wires with length of 30 $\mu$m), fabricated with home-made AAM, from room temperature down to 0.47 K. The dashed line shows that of the 100 nm Sn wires (53 wires in the array with length of 6 $\mu$m), fabricated with PCM. The resistance of the 100 nm wires is normalized to the value of the 20 nm wires at 5 K by a multiplicative factor of 230. The R(T) curves of these two samples show metallic behavior from room temperature down to $T_c$ with a room temperature to 5 K residual resistance ratio (RRR) of 9.5 ± 0.5. This means, as we have indicated above, that our squeezing technique has broken through the oxide layers and that we have established direct contact bulk Sn to the nanowires. Indeed in squeezing the Sn leads onto the nanowire arrays, we made sure the room temperature resistance value is consistent with that without an oxide barrier between the leads and the nanowires. Since rather high pressures are required to made good, i.e. oxide barrier free, contacts, one may be concerned that this technique in establishing contact

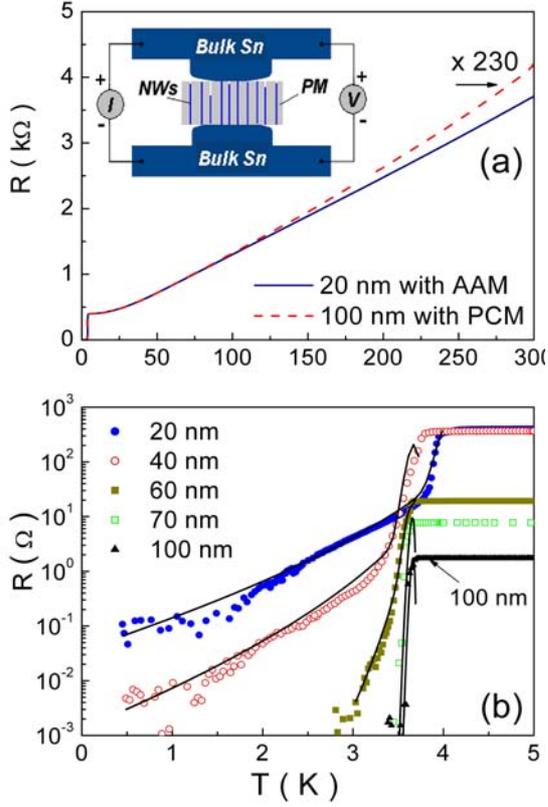

Fig. 2 (a) Resistance versus temperature of 20 and 100 nm Sn wires fabricated with AAM and PCM, respectively, in the wide temperature range of 0.47-300 K. The resistance value of 100 nm wires was normalized to that of 20 nm wires at 5 K by multiplicative a factor of 230. A schematic arrangement for the transport measurement is shown in the inset. Fig.2(b) summarizes the results of our R(T) measurement in low temperature range for the 20, 40, 60, 70 and 100 nm Sn nanowire arrays. The solid lines for 20, 40 and 60 nm wires are the calculation results based on TAPS model near $T_c$ and QPS model below $T_c$ with four adjusting parameters, while those for 70 and 100 nm wires were made just based on TAPS model with two fitting parameters.

may induce defects in the nanowires. The similarity of the R vs T curves shown in Fig. 2(a) suggests this is not the case. The RRR of the 20 nm wires fabricated in AAM and the 100 nm wires fabricated in PCM are found to be nearly identical, in spite of the fact that the Young's modulus of AAM at 122 GPa[16] is nearly 50 times larger than that of PCM[17]. It appears the membranes are effective in protecting the structural integrity of the nanowires inside the pores. The RRR of our single-crystal wires at 9 is 4 to 5 times larger than polycrystalline Zn and Au wires of comparable diameter fabricated and measured with similar techniques. A RRR ratio of 9, however is an order of magnitude smaller than those of indium, tin, and lead whiskers[8-10]. There are a number of possible reasons for the reduced RRR. These include the intrinsic limitation of the crystal quality of the wires in the fabrication process (like twin boundaries observed in TEM studies), enhanced surface scattering due to the small diameter of the wires and strains in the wires induced by differential thermal contraction between the wires and the membrane upon cooling.

Figure 2(b) shows the as-measured resistance of the bulk Sn/Sn-nanowires/bulk Sn system as a function of temperature for wires of different diameters. The excitation dc currents used for 20, 40, 60 nm wire array are 1.0 $\mu$A, while those for 70 and 100 nm thick wire array are 10.0 $\mu$A. The current densities in 20, 40, 60, 70, and 100 nm samples are estimated to be, respectively, $1.8\times10^4 A/cm^2$, $8.0\times10^4 A/cm^2$, $0.44\times10^4 A/cm^2$, $1.7\times10^4 A/cm^2$, and $0.56\times10^4 A/cm^2$. These excitation currents are well below the critical current value ($I_c$) of the nanowires (details on $I_c$ will be discussed below). Since the series bulk Sn leads and the contact resistance are negligible below $T_c$ we are able to follow the resistance of the nanowire down to the resolution limit of our measurement system at about $\sim 5\times10^{-3}$ $\Omega$ with an excitation current of 1.0 $\mu$A. The onset temperature $T_c$ of nanowires, i.e., the temperature at which resistance shows an abrupt drop, was found to be close to the bulk value at 3.7 K for all wires with $d \geq 40$ nm. A slight increase in $T_c$ to 4.1 K is found for the thinnest, i.e. 20 nm wire. It should be noted that the value of $T_c$ for both the thicker and the 20 nm wires are reproducible to better than 0.1 K for wires fabricated at different times. Indeed all the key experimental features reported in this paper have been checked for their reproducibility with multiple samples. The behavior of $T_c$ of our single-crystal wires is different from that of granular wires and amorphous wires. In granular wires[18,19] and granular films[20], the superconducting transition temperature $T_c$ is generally found to be higher than that of bulk $T_c$, but insensitive to the cross-section area of wires. On the other hand, $T_c$ of amorphous wires was found to be significantly suppressed with decreasing wire cross-section area. The enhancement of $T_c$ found in the 20 nm single-crystal Sn wires may have the same origin of that of granular films and that of granular wires. One possible origin is that this is an effect of the increased surface area, which may enhance surface electron-phonon scattering effects[21].

The resistance of wires with diameters larger than 70 nm drops precipitately to zero within 0.2 K of $T_c$, but the transition of 60 nm wires broadens and begins to show a residual resistance extending down to 3.0 K with the resistance falling below $10^{-3}$ $\Omega$, the resolution of our measurement. In wires of 40 and 20 nm, a clear functional dependence of the residual resistance on temperature is found. Specifically, when the logarithm of the resistance is plotted vs. temperature, two distinct linear sections are found. A high temperature linear section is found immediately below $T_c$. Another linear section extends from just a few tenths of a degree below $T_c$ down to the lowest temperature of measurement at 0.47 K. The broadening of the transition near $T_c$ in thin whiskers[22,23] and In nanowires[1] has been observed and was interpreted as a consequence of a thermally activated phase slip (TAPS) process[24]. The TAPS model predicts a dissipative resistance $R_{TAPS}$ that scales as $e^{-\Delta F/k_B T}$, with $\Delta F = (8\sqrt{2}/3)(H_c^2/8\pi)A\xi$ being the free energy barrier, and $H_c$, $\xi$, A, $k_B$ being the temperature dependent critical filed of bulk sample, phase coherence length, cross-sectional area of 1d wire and the Boltzmann constant, respectively. Since $H_c$ goes as $(T_c-T)$, and $\xi \sim (T_c-T)^{3/2}$ near $T_c$, the energy barrier $\Delta F \sim (T_c-T)^{3/2}$. This means the term, $e^{-\Delta F/k_B T}$ drops off very rapidly when the temperature is decreased from $T_c$. Therefore, the TAPS model is expected to be relevant only at temperature very close to $T_c$. The exponential residual resistance near $T_c$ as shown in Fig. 2(b) for 20, 40 and 60 nm is consistent with these earlier experiments.

The second low temperature exponential residual resistance that extends over a wide range of temperature down to 0.47 K cannot be understood in the framework of the TAPS model. A similar exponential decay in resistance down to low temperature was found by Giordano[1] in granular indium wire of 41 nm in diameter, fabricated from evaporated In fillm with step-edge lithographic technique[1,25]. In Giordano's experiment, the low temperature exponential decay in resistance is found to extend from $T_c = 3.4$ K down to 2.7 K, the lowest temperature of his experiment. Giordano proposed a phenomenological quantum phase slip (QPS) model to explain his results. The finite exponentially decaying in resistance is proposed to result from quantum fluctuation induced tunneling through an energy barrier $\Delta F$, resulting in a resistance $R_{QPS}$ of the form, $e^{-\Delta F \tau_{GL}/\hbar}$ with $\tau_{GL}$ and $\hbar$ being the Ginzburg-Laudau relaxation time and Plank's constant. The solid lines tracing the measured resistance of the 20, 40 and 60 nm below $T_c$ are fits of the form $R_{total} = R_{TAPS} + R_{QPS}$ [26]. The $R_{TAPS}$ and $R_{QPS}$, respectively, denote the contribution of resistances from TAPS process near T and QPS process at temperature below $T_c$. The fact that we are able to fit to the data well with four free parameters over the full temperature range lends support to the proposal of Giordano of a QPS process.

In order to further understand these interesting transport properties in thin single-crystal Sn nanowires, the voltage-current (V-I) characteristics were measured at different temperatures. The results of 70, 40 and 20 nm Sn wires are shown in Fig 3. We shall first explain our results in the low excitation current more clearly displayed in the log-log scale (panels d, e and f). The 70 nm wires (panel d) at any temperatures of $T < T_c$ show exactly zero voltage when the excitation current is below a certain temperature dependent critical value $I_c$. In contrast, the 20 nm nanowire array (panel f), even at low current limit, never shows zero voltage at all temperatures. The V-I curves are parallel to each other in the current range $I < 2$ $\mu$A and also parallel to that in the normal state at T=4.2 K. This demonstrates that the finite resistance measured at low current limit ( $I < 2$ $\mu$A ) in the thinner wires shown in Fig.2 (b) is ohmic in nature, not only near $T_c$, but also at temperatures well below the $T_c$. Similar low current ohmic resistance behavior was also found in the 40 nm individual wire, but the measured voltage becomes vanishing small at temperatures below 2.5 K. The fact that the ohmic finite resistance found in the low excitation current limit is enhanced with reduced diameters suggests this is likely an effect due to 1d confinement. A quantitative prediction of the TAPS model is that the residual resistance measured in the low current limit should be ohmic[22,23]. This is a consequence that in the TAPS model, the residual resistance results from an activated phase-slip process over an energy barrier. Our I-V measurements here show that this is in fact the case for the residual resistance for T near $T_c$. Furthermore, we found the residual resistance of the 20 and 40 nm wires to be ohmic at all temperatures. This lends further credence to the phenomenological QPS model with a tunneling energy barrier.

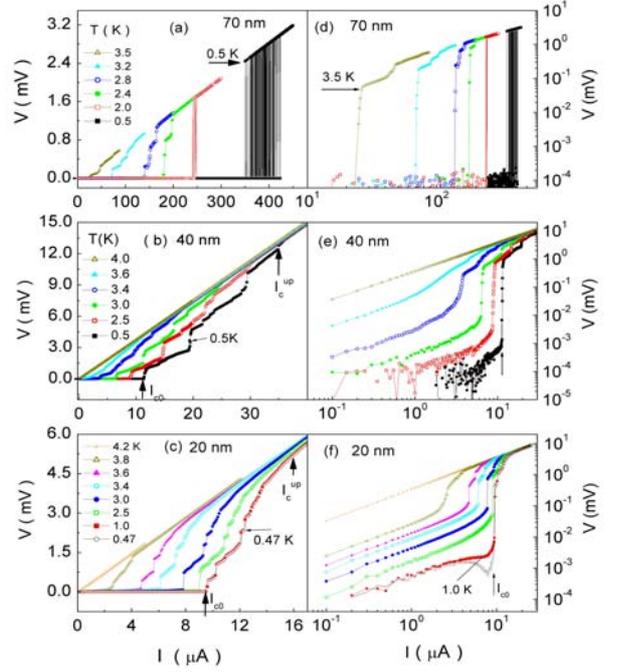

Fig. 3 V-I curves of (a) 70, (b) 40 and (c) 20 nm Sn nanowire arrays measured at different temperatures in linear scale. Panels (d), (e) and (f) show log-log plots of V versus I. These plots show more clearly the data in the low current limit.

When the excitation current in 20 and 40 nm wires is increased beyond the linear region, as seen in the panel (e) and (f), the measured voltage first shows an upward deviation and then jumps up sharply with current. The value of the current at the initial point of the voltage jump in the V-I curve is defined as the initial critical current, $I_{c0}$. They are, respectively, about 9.5 and 11.0 $\mu$A at 0.47 K for 20 and 40 nm wires, which corresponds to current densities, $j_c$ of $1.8 \times 10^5$ A/cm$^2$ and $8.8 \times 10^4$ A/cm$^2$. These values are about 1-2 orders smaller than that for bulk Sn[27], $j_c \sim 2.3 \times 10^7$ A/cm$^2$. As shown in panel (f), there is a minimum in voltage just below the $I_{c0}$ in the 20 nm wires for temperature below 1.0 K. We do not understand the mechanism of this interesting behavior.

The V-I curves of 20, 40 and 70 nm wires in high excitation current are more clearly displayed in linear scales (panels a, b, and c in Fig.3). The behavior found for the 70 nm wires is again qualitatively different from that of 20 nm and 40 nm wires. The V-I curves for 20 and 40 nm wires showed a series of voltage steps in approaching the normal state over the entire temperature range of $T<T_c$. The current value at which the wires are driven to normal state is defined as the upper critical current $I^{up}_c$. These voltage steps between $I_{c0}$ and $I^{up}_c$ are reproducible irrespective of whether the current is swept up or down. However, the steps are not found in the 70 nm wires for T<2.4 K, instead, upon reaching a certain excitation current, the voltage fluctuates between zero in the superconducting state and a finite value in the normal state. Upon further increase of the excitation current, the measured voltage stabilized in the normal state. There are no two distinct critical currents for 70 nm wires as in 20 and 40 nm wires at temperature below 2.4 K. A likely explanation of this fluctuation behavior is that when one or more of the wires are driven to the normal state due to the dispersion of wire diameters, the high finite resistance causes heating and induces the entire wire array to the normal state. In the normal state the higher resistance limits the flow of the current through the wires, thus allowing the wire array to cool and return to the superconducting state. This process repeats over a range of the excitation current. One step in the V-I curve emerges between the superconducting and the normal state in the 70 nm wires as in 20 and 40 nm wires at 2.4 K and multiple steps are clearly seen at 3.2 and 3.5 K. These results suggest that the voltage steps are consequence of the wire approaching the 1d limit. Near $T_c$ the superconducting coherence length increases, which has the effect of placing the 70 nm wire closer to the 1d limit. On the other hand, wires of 40 nm and thinner diameters appear to be in the 1d limit over the whole temperature range. It is noteworthy that the exponential residual resistances, measured under low excitation current as shown in Fig.2 (b), are also seen in 40 and 20 nm wires but not in wires thicker than 60 nm.

The voltage steps seen in Sn nanowires are reminiscent of those observed previously in Sn whiskers[7-9,28], but the steps in whiskers can be seen only in a very narrow temperature region ($\Delta T \sim 0.1$ K) below the $T_c$. They were interpreted as a consequence of spatially localized "weak spots" r phase slip centers (PSCs)[29,30]. These "weak spots" or PSCs were thought to be related to the local defects or imperfections in the whisker with a smaller local critical current. When the excitation current exceeds the local critical current of a specific PSC, a step in the whisker is found. The spatial extend of the PSC is expected to be on the order of several micrometers, determined by the quasi-particle diffusion length, $\Lambda_q$, due to the conversion of normal electrons into Cooper pairs near the PSC. The number of voltage steps in V-I curves was found to scale with the length of the whisker[7]. Our results show several features that are different from that found in whiskers and cannot be explained by the classical Skocpol-Beasley-Tinkham (SBT) PSC model [29,30]. The first is the number of steps we found in the 40 nm and 20 nm wires. Since these wires were fabricated with identical procedures, they should be similar in terms of crystallinity and imperfection. If the line density of the weak spots is similar, then the 20 nm wire array with 18 wires of 30 $\mu$m in length should have 90 times more PSCs than that in the single 40 nm wire of 6 $\mu$m in length. Instead, Fig.3 shows a comparable number of sharp steps. Secondly, we did not observed the hysteresis in the V-I curves near the voltage steps as expected by the theoretical predictions[31]. Thirdly, the step number was found to decrease with increasing the magnetic field, as shown in Fig.4 (by increasing the magnetic field beyond a certain value, 3.0 kOe at 2.5 K, the multiple voltage steps merge into two and then a single step). This is in contrast to the prediction of the SBT model, in which the major effect of increasing H is the decrease of the differential resistance in the plateau region of the V-I curves, together with the corresponding increase in the number of PSCs that can "fit" along the length of the wire (the $\Lambda_q$ was found to decrease with increasing H [30]).

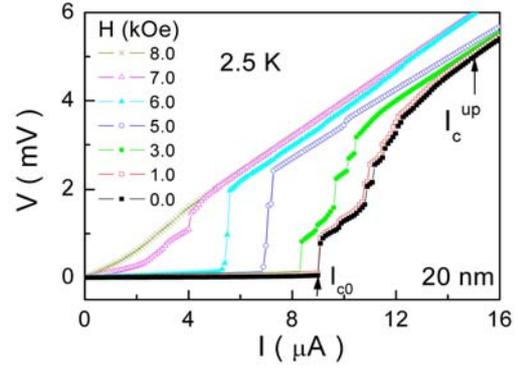

Fig. 4 V-I curves of 20 nm Sn wires measured at 2.5 K and different magnetic fields aligned parallel to the wires.

Very recently, Michotte et al [12,32,33] investigated the V-I characteristics of polycrystalline Sn and Pb nanowires, fabricated by electrodepositing Sn and Pb into porous polycarbonate membranes. They observed two voltage steps in the V-I curve at temperature far from $T_c$ under constant current-driving mode measured with two Au-leads. Except for some differences in details between our data and Michotte's, the overall behavior of their V-I characteristic of Sn wires is, qualitatively, similar to ours, including no hysteresis in V-I curves under constant current-driving mode, and the disappearance of the voltage steps with the increase of magnetic field H. It is noteworthy that the length and diameter of the Sn wires in Michotte's work are respectively 50 $\mu$m and 55 nm, 8 times longer than our 40 nm individual wire in length. The extend of one PSC in the wire is estimated to be around 40 $\mu$m[12], also 6 times longer than the total length of our 40 nm sample. In order to understand the data seen in Sn and Pb nanowires, Michotte et al [32,33] extended the classical PSC model to 1d nanowires by considering two different boundary conditions (bridge (S-S) and N-S) using the generalized time-dependent Ginzburg-Landau (TDGL) equation. They considered the effect of the defects (a local variation of the critical temperature and a local variation of the cross-section of the wire) and magnetic field on the V-I characteristic. The important conclusions in their theory are that: (1) with applying a magnetic field, there exist a critical field $H^*$ above which there will be no voltage jump (PSC) in the V-I curves. This is because the magnetic field suppresses the order parameter, $|\psi|$, everywhere in the sample, it thus leads to an increase of $\Lambda_q \sim 1/\sqrt{|\psi|}$, in TDGL model and hence there is a lack in space for the coexistence of two more PSCs in the wire at H>$H^*$ (in contrast to SBT model). (2) hysteresis in the V-I curves will disappear when the defect is sufficiently "strong". These predictions are consistent with our observation in our V-I characteristics. However, Michotte's theory also predicted that there exists a critical length $L^*$ below which there will be no PSC. The theory still cannot explain why our relatively short 40 nm individual wire show more sharp steps and the step number of the V-I curves in our single-crystal Sn wires is insensitive to the length and the number of the wires in the array. Our findings suggest that the voltage steps due to the PSCs found in our nanowire array is not merely determined by the independent local defects or weak spots in each wire, but also involve competition mechanisms among these PSCs.

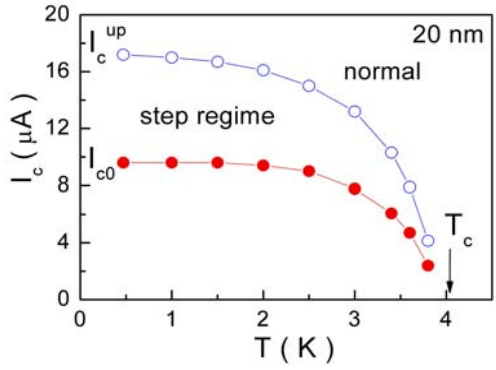

Fig. 5 $I_c$-T plot for 20 nm Sn wires.

Fig.3 shows different V-I behavior in 20 and 40 nm wires for excitation current below and above the initial critical current $I_{c0}$. When the current is higher than the upper critical value $I^{up}_c$, the wires become completely normal. Fig.5 shows a plot of $I_{c0}$ and $I^{up}_c$ as a function of temperature showing two regimes with different dissipation processes. At excitation currents below $I_{c0}$, the nanowire system is in the homogenous superconducting state. The dissipations are governed by TAPS or QPS processes. This regime is separated from the normal state by a regime characterized by multiple voltage steps, in which the nanowires are driven into current-induced inhomogeneous strong dissipation state (i.e., PSC state).

In summary, we found the electrical transport properties of single-crystal Sn wires with diameters of equal and below 40 nm to be distinctly different from that of thicker wires. This suggests Sn wires with diameters less than 40 nm, which is 5 times smaller than the phase coherence length, has reach the 1d limit. The 1d wires show evidence of quantum fluctuation induced dissipation in the low temperature regime under low excitation current. When the excitation current exceeds a critical value, the voltage-current (V-I) curves show a series of discrete steps in approaching the normal state. The mechanism of these multi-steps in this 1d nanowire array is not understood.

**Acknowledgments** We acknowledge fruitful discussions with V. Crespi, N. Giordano, J. Han, J. Jain, P. A. Lee, P. Schiffer, M. Tinkham, and X. G. Wen. This work is supported by Penn State MRSEC funded by NSF under grant DMR-0213623.